\documentclass[journal]{IEEEtran}

\usepackage{cite}
\ifCLASSINFOpdf
  \usepackage[pdftex]{graphicx}
  \graphicspath{{../img/}}
  \DeclareGraphicsExtensions{.pdf,.jpeg,.png}
\fi
\usepackage{url}

\begin{document}

\title{Open storm: a complete framework for sensing and control of urban watersheds}%

\author{Matthew~Bartos,~Brandon~Wong,~and~Branko~Kerkez
\thanks{M. Bartos, B. Wong and B. Kerkez are with the Department
of Civil and Environmental Engineering, University of Michigan, Ann Arbor
MI, 48105 USA}
\thanks{Electronic Supplementary Information (ESI) available: github.com/open-storm/docs.open-storm.org}}

\markboth{ArXiv Pre-Print}%
{Bartos \MakeLowercase{\textit{et al.}}: Open storm: a complete framework for sensing and control of urban watersheds}
\maketitle

\begin{abstract}
  Leveraging recent advances in technologies surrounding the \textit{Internet of
    Things}, ``smart'' water systems are poised to transform water resources
  management by enabling ubiquitous real-time sensing and control. Recent
  applications have demonstrated the potential to improve flood forecasting,
  enhance rainwater harvesting, and prevent combined sewer overflows. However,
  adoption of smart water systems has been hindered by a limited number of
  proven case studies, along with a lack of guidance on how smart water systems
  should be built. To this end, we review existing solutions, and introduce
  \textit{open storm}---an open-source, end-to-end platform for real-time
  monitoring and control of watersheds. \textit{Open storm} includes (i) a
  robust hardware stack for distributed sensing and control in harsh
  environments (ii) a cloud services platform that enables system-level
  supervision and coordination of water assets, and (iii) a comprehensive,
  web-based ``how-to'' guide, available on \texttt{open-storm.org}, that
  empowers newcomers to develop and deploy their own smart water networks. We
  illustrate the capabilities of the \textit{open storm} platform through two
  ongoing deployments: (i) a high-resolution flash-flood monitoring network that
  detects and communicates flood hazards at the level of individual roadways and
  (ii) a real-time stormwater control network that actively modulates discharges
  from stormwater facilities to improve water quality and reduce stream erosion.
  Through these case studies, we demonstrate the real-world potential for smart
  water systems to enable sustainable management of water resources.
\end{abstract}

\section{Introduction}

\IEEEPARstart{A}{dvances} in wireless communications and low-power sensing are
enabling a new generation of ``smart cities,'' which promise to improve the
performance of municipal services and reduce operating costs through real-time
analytics and control \cite{caragliu_2011}. While some applications of ``smart''
infrastructure have received a great deal of attention---such as autonomous
vehicles\cite{Dimitrakopoulos_2010, Zanella_2014}, energy grid management
\cite{Zanella_2014}, and structural health monitoring\cite{lynch_2005,
  Zanella_2014}---integration of these technologies into water systems has
lagged behind. However, ``smart'' water systems offer new inroads for dealing
with many of our most pressing urban water challenges, including flash flooding,
aquatic ecosystem degradation, and runoff pollution. The goal of this paper is
to provide an end-to-end blueprint for the next generation of autonomous water
systems, with a particular focus on managing urban stormwater. Towards this
goal, we introduce \textit{open storm}, an open source framework that combines
sensing, real-time control, wireless communications, web-services and
domain-specific models. We illustrate the potential of \textit{open storm}
through two real-world case studies: 1) a 2,200 km$^2$ wireless flood
forecasting network in Texas, and 2) an 11 km$^2$ real-time stormwater control
network in Michigan. Most importantly, to encourage broader adoption by the
water resources community, this paper is accompanied by extensive supplementary
materials on \texttt{open-storm.org}, including videos, photos, source code,
hardware schematics, manufacturing guides, and deployment instructions. These
materials make it possible for newcomers to implement their own ``smart''
stormwater systems, without extensive experience in programming or embedded
systems design.

\section{Background}

\begin{figure*}[!ht]
\centering
\includegraphics[width=\textwidth]{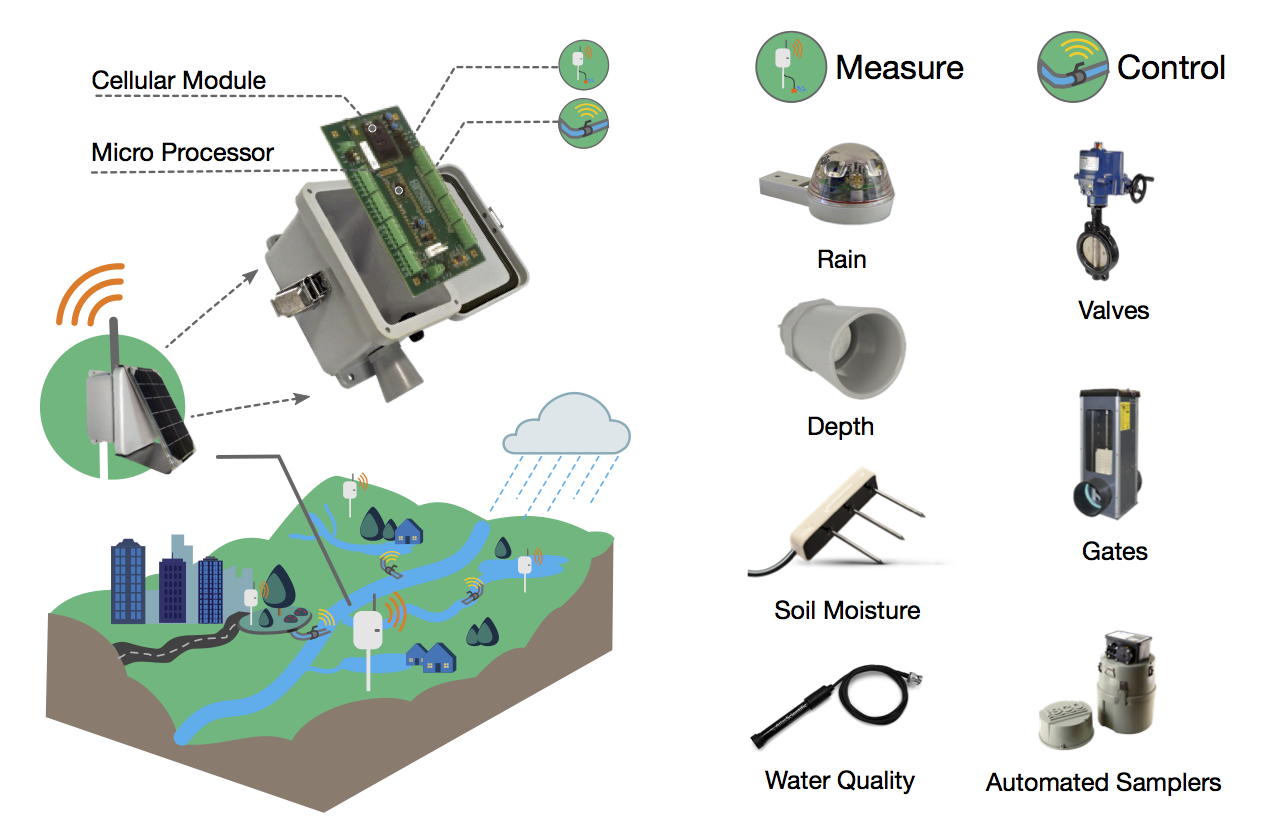}
\caption{The \textit{open storm} hardware layer. The left panel shows the
  complete sensor node along with a representative schematic of its placement in
  an urban watershed. The right panel shows typical sensors and actuators used
  in \textit{open storm} research projects.}
\label{fig:1}
\end{figure*}

\subsection{Motivation}

Effective management of water supply and water excess are some of the largest
engineering problems faced by cities today \cite{mays_2010}, and in the wake of
rapid urbanization, aging infrastructure, and a changing climate, these
challenges are expected to intensify in the decades to come
\cite{bronstert_2002, stocker_2014}. Floods are the leading cause of severe
weather fatalities worldwide, accounting for roughly 540,000 deaths between 1980
and 2009 \cite{doocy_2013}. Furthermore, large quantities of metals, nutrients,
and other pollutants are released during storm events, making their way via
streams and rivers into lakes and coastal zones \cite{Ahn_2005, Carey_2014}. The
need to manage pollutant loads in stormwater has persistently been identified as
one of our greatest environmental challenges \cite{V_r_smarty_2010}. To contend
with these concerns, most communities maintain dedicated gray infrastructure
(pipes, ponds, basins, wetlands, etc.) to convey and treat water during storm
events. However, many of these systems are approaching the end of their design
life \cite{epa_2016}. At the same time, stormwater systems are being placed
under greater stress due to larger urban populations, changes in land use, and
the increasing frequency of extreme weather events\cite{mays_2010,
  stocker_2014}. In some communities, stormwater and wastewater are combined,
meaning they share the same pipes. For these systems, large storms often lead to
combined sewer overflows, which release viruses, bacteria, nutrients,
pharmaceuticals, and other pollutants into estuaries downstream
\cite{Sercu_2011}. When coupled with population stressors, it comes as little
surprise that the current state of stormwater infrastructure in the United
States has been given a near-failing grade by the American Society of Civil
Engineers \cite{asce_2013}.

Engineers have traditionally responded to increasing demands on stormwater
systems by expanding and constructing new \textit{gray} infrastructure. However,
the upsizing of pipes and storage elements can prove expensive, time-consuming,
and may even result in deleterious long-term side effects. Benefits from
stormwater conveyance facilities can be diminished if individual sites are not
designed in a global context. Even when best management practices are followed,
discharges from individual sites may combine to induce downstream flows that
are more intense than those produced under unregulated conditions \cite{Emerson_2005}.
Without system-level coordination, gray infrastructure expansion may lead to
overdesigned solutions that adversely impact flooding, increase stream erosion,
and impair water quality\cite{Hawley_2016}. In response to these concerns,
\textit{green} infrastructure (GI) has been proposed as an alternative to
traditional ``steel and concrete'' stormwater solutions. These systems use
smaller, distributed assets---such as bioswales, green roofs and rain
gardens---to condition flows and improve water quality. However, recent research
has raised questions about the scalability and maintenance requirements of green
infrastructure \cite{epa_2013}. Regardless of the choice between ``gray'' or
``green'', new construction is limited by cost, and often cannot keep pace with
evolving community needs. To preserve watershed and ecological stability, there
is an urgent need to incorporate systems thinking into stormwater designs and to
engineer solutions that can optimize stormwater system performance---not only
for individual sites, but for entire watersheds.

\subsection{The promise of sensing and control}

``Smart'' water systems promise to improve the study and management of water
resources by extending monitoring and control beyond centralized facilities and
into watersheds as a whole. With increased access to inexpensive sensors and
wireless communications, the feasibility of deploying and maintaining large
sensor networks across urban landscapes is now within reach for many public
utilities and research groups. While many of the technologies have existed for
some time, it was not until the integration of wireless sensor networks with web
services (i.e. the \textit{Internet of Things}) that large networks consisting of
hundreds or thousands of heterogeneous devices could be managed
reliably\cite{wong_2016b}. This in turn has enabled watersheds to be studied at
spatial and temporal scales that were previously unattainable. By densely
instrumenting urban watersheds, researchers can finally begin to understand the
complex and spatially variable feedbacks that govern water flow and quality
across the built environment. A system-level understanding of urban watershed
dynamics will provide decision makers with actionable insights to alert the
public, and improve stewardship of water resources.

Beyond new insight gained through sensing, the ability to dynamically regulate
water levels across a watershed will reduce flooding, preserve riparian
ecosystems, and allow for distributed treatment of stormwater. While these
functions were previously achieved only through construction of static gray
infrastructure or centralized treatment facilities, the addition of
remotely-controlled valves and pumps promises to realize the same benefits while
at the same time reducing costs, expanding coverage, and allowing system
performance to scale flexibly with changing hydrologic conditions. Adding valves
to existing stormwater facilities, for instance, can extend hydraulic retention
time, thereby promoting the capture of sediment-bound
pollutants\cite{kerkez_2016, Mullapudi_2017}. Modulation of flows (hydrograph
shaping) may reduce erosion at downstream locations by ensuring that discharges
do not exceed critical levels\cite{kerkez_2016}. More fundamentally, distributed
control will enable operators to coordinate outflows from stormwater sites (tens
to hundreds) across an entire city. Along with reducing flooding, this will
allow water managers to utilize the latent treatment capacity of existing ponds
and wetlands---effectively allowing a watershed to function as a distributed
wastewater treatment plant\cite{Mullapudi_2017}.
 
Such a vision for ``smart'' stormwater systems is no longer limited by
technology. Rather, adoption of smart water systems has been hindered by (i) a
reliance on proprietary technologies, (ii) a lack of proven case studies, and
(iii) an absence of end-to-end solutions that are specifically designed and
tested for water resources applications. To enable truly holistic management and
control, there is an urgent need to combine modern technologies with domain
knowledge from water sciences---something which present solutions do
not address or make transparent. These solutions are reviewed next, after which
the \textit{open storm} framework is introduced as an end-to-end blueprint for
``smart'' water systems management. This open-source framework combines
low-power wireless hardware with modern cloud computing services and
domain-specific applications to enable scalable, real-time control of urban
water systems.

\section{Existing technologies}

Real-time sensing and control of water infrastructure is not a new idea.
Supervisory control and data acquisition (SCADA) systems have long been used to
monitor and control critical water infrastructure \cite{mays_2000}. In addition
to these traditional technologies, there has been a recent explosion in the
development of wireless sensor networks (WSNs) for water resources management.
While these technologies have made great strides in enabling monitoring and control
of water systems, a lack of end-to-end solutions has inhibited system-scale
management of watersheds. In this section, we review existing technological
solutions for water system monitoring and control, and describe how \textit{open
  storm} advances the state of the art by providing the first open source,
end-to-end solution for distributed water systems management.

\subsection{SCADA systems}

Most water utilities use supervisory control and data acquisition (SCADA)
systems to manage the conveyance, treatment and distribution of water
\cite{mays_2000}. These systems comprise collections of devices, communication
protocols, and software that enable remote monitoring and control of water
assets \cite{mays_2000}. Most commonly applied in water distribution systems,
SCADA systems typically monitor parameters that indicate service quality---such
as flows, pressures, and chemical concentrations---and then use this information
to control the operation of pumps and valves in real-time \cite{mays_2000}.
Control may be manual or automatic, and in some cases may integrate optimization
algorithms, decision support systems and advanced control logic
\cite{mays_2000}. While legacy SCADA systems remain popular among water
utilities, they suffer from limitations in three major areas: interoperability,
scalability and security.

Perhaps the most critical limitation of legacy SCADA systems is the lack of
interoperability between systems, reliance on proprietary protocols, and
non-extensible software \cite{powell_1999}. Traditional SCADA systems are often
isolated and incapable of intercommunication \cite{powell_1999}. Systems that
manage water in one municipality, for instance, may be incapable of
communicating with those in another municipality, despite sharing the same
service area. Moreover, different SCADA systems within the same jurisdiction may
also be isolated, meaning that management of stormwater systems may not in any
way inform the operation of wastewater treatment facilities downstream. This
lack of communication between water management architectures makes it difficult
to coordinate control actions at the watershed scale. Proprietary SCADA systems
are also often unable to interface with modern software layers, like Geographic
Information Systems (GIS), network analysis software, or hydrologic models
\cite{powell_1999}. For this reason, SCADA-based control often cannot take
advantage of modern domain-specific tools that would enable system-scale
optimization of watershed resources.

The capacity of SCADA systems to implement watershed-scale control is also
limited by a lack of spatial coverage. Due to their large power footprint and
maintenance requirements, traditional SCADA systems are typically limited to
centralized water assets with dedicated line power, such as drinking water
distribution systems and wastewater treatment facilities \cite{awwa_2001}.
Sensors are usually deployed at a select few locations within the network---like
treatment plants, pump stations and boundaries with other systems---and in many
cases plant and pump station discharges are not even recorded \cite{mays_2000}.
For decentralized applications, such as stormwater networks or natural river
systems, the cost and power usage of traditional SCADA systems are prohibitive.
As such, these distributed resources often go unmonitored and uncontrolled.

Recent studies have also raised concerns about the security of SCADA systems,
many of which were designed and installed decades ago
\cite{igure_2006,mays_2004}. Many legacy SCADA systems rely on specialized
protocols without built-in support for authentication, such as MODBUS/TCP,
EtherNet/IP and DNP317 \cite{igure_2006,mays_2004}. The use of unsecured
protocols means that it is possible for unauthorized parties to execute commands
remotely on a device in the SCADA network \cite{igure_2006}. To cope with this
problem, SCADA networks are often isolated from public networks, such as the
internet. However, remote attacks are still possible---particularly through the
use of unsecured radio channels \cite{mays_2004}. Moreover, isolation from
public networks limits the use of modern web services such as cloud computing
platforms. Reliance on closed networks and proprietary interfaces may also lend
a false sense of security to legacy SCADA systems---a concept known as security
through obscurity \cite{igure_2006}. For these reasons, SCADA systems have
gained the reputation of being relatively closed and only manageable by
highly-trained operators or specialized local consultants. While SCADA systems
remain the most popular platform for managing urban water systems, new tools are
needed to improve security, expand coverage, and encourage integration with
modern software.

\subsection{Wireless sensor networks}

The past decade has witnessed a large reduction in the cost and power
consumption of wireless electronics; leveraging these advances, wireless sensor
networks (WSNs) have opened up new frontiers in environmental monitoring, with
applications ranging from biodiversity monitoring \cite{cerpa_2001}, forest fire
detection \cite{hefeeda_2007, soliman_2010}, precision agriculture
\cite{kim_2008}, glacier research \cite{martinez_2004}, and structural health
monitoring \cite{lynch_2005}. Unlike SCADA systems, WSNs are ideal for low-cost,
low-power, and low-maintenance applications, making them well-suited for the
monitoring of large water systems like rivers and watersheds. WSNs have been
applied to great success in applications ranging from flood monitoring to
real-time stormwater control; however, current implementations are
generally experimental or proprietary, resulting in a lack of discoverability,
limited interoperability, and duplication of effort among projects.

Within the water sciences, flood monitoring represents a particularly important
application area for WSNs. While several groups have worked to expand the
capabilities of existing legacy flood detection networks \cite{bonnet_2000,
  imielinski_2000, chen_2014}, only a small number of groups have designed and
deployed their own flood monitoring WSNs. Hughes et al. (2008) describe a
15-node riverine flood monitoring WSN in the United Kingdom, which interfaces
with remote models, performs on-site computation, and sends location-specific
flood warnings to stakeholders \cite{hughes_2008, smith_2009}. Other riverine
flood monitoring networks include a 3-node river monitoring network in
Massachusetts, a 4-node network in Honduras \cite{basha_2008}, and---perhaps the
largest unified flood monitoring network in the US---the Iowa Flood Information
System (IFIS), which draws on a network of over 200 cellular-enabled sensor
nodes \cite{demir_2013}. While most existing flood-monitoring networks focus on
large-scale river basins, flash-flooding has received considerably less
attention in the WSN community. Marin-Perez et al. (2012) construct a 9-node WSN
for flash flood monitoring in a 660 km$^2$ semiarid watershed in Spain
\cite{marin_perez_2012}, while See et al. (2011) use a Zigbee-based WSN to
monitor gully-pot overflows in an urban sewer system \cite{see_2011}. While most
deployments are still pilot-scale, these projects demonstrate the potential of
WSNs for distributed flood monitoring across a variety of scales and
environments.

In addition to monitoring watershed hazards, a limited---but promising---number
of projects are illustrating the potential of WSNs for real-time control.
Web-enabled sensor nodes have been used to develop adaptive green infrastructure
at a select number of pilot sites---for instance, by using weather forecasts to
facilitate predictive rainwater harvesting and capture of sediment-bound
pollutants\cite{ewri_2015}. At larger scales, a combined sewer network in South
Bend, Indiana uses over 120 flow and depth sensors along with nine valves to
actively modulate flows into the city's combined sewer system
\cite{montestruque_2015}. This network optimizes the use of existing in-line
storage and has achieved a roughly five-fold reduction in combined sewer
overflows from 2006-2014 \cite{montestruque_2015}---all without the construction
of additional infrastructure. While distributed control of storm and sewer
systems shows promise, most existing implementations are proprietary. A lack of
transparency makes these solutions inaccessible to decision makers and the water
resources community at large.

Although many research groups have realized the potential for real-time
watershed monitoring, existing WSN deployments are generally small-scale and
experimental in nature. In order for these networks to be accepted as ``critical
infrastructure'' by the water resources community at large, consistent standards
for design, deployment and functionality are needed. In designing their own
WSNs, researchers tend to look towards previous research projects
\cite{basha_2008}. However, research papers rarely include the detailed
documentation needed to implement an end-to-end sensor platform
\cite{basha_2008}. As a result, researchers are often forced to design and
deploy their own WSNs from scratch. To prevent duplication of effort and ensure
best practices, a community-driven \textit{how-to guide} is urgently needed.
Moreover, while proprietary control networks have proven their effectiveness in
improving the performance of stormwater systems, an open source alternative is
needed to encourage transparency, interoperability, and extensibility. Without
open software, standards, and documentation, these new technologies risk
becoming like the SCADA systems of old: isolated, proprietary, and incapable of
intercommunication.


\begin{figure*}[!htb]
\centering
\includegraphics[width=\textwidth]{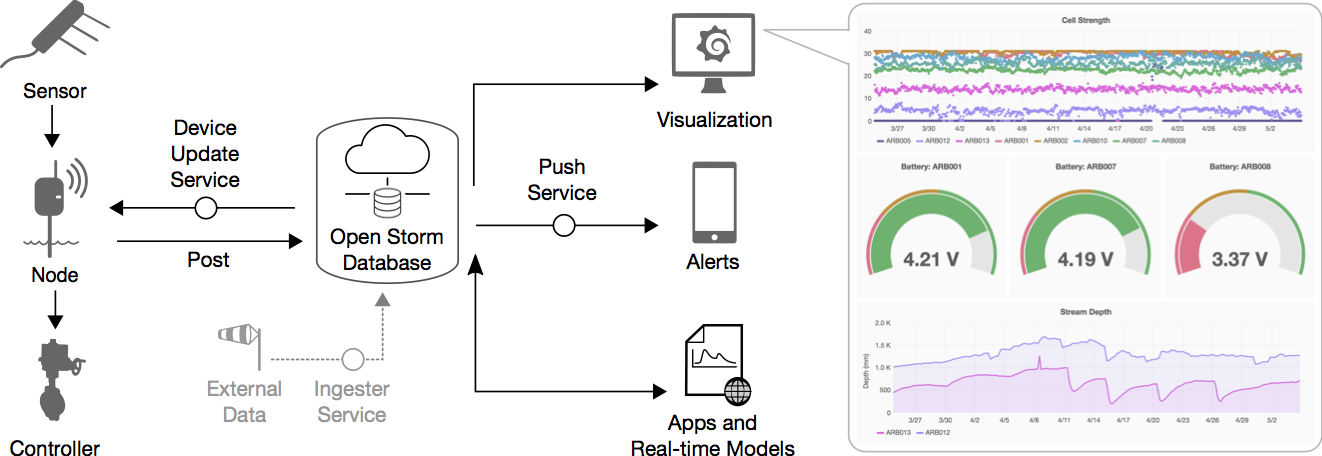}
\caption{The \textit{open storm} stack. The hardware layer (left) comprises the sensor
  node along with auxiliary sensors and actuators. The cloud services layer
  (center) includes the database backend, along with a series of publication and
  subscription services for controlling sensor node behavior and interfacing
  with applications. The applications layer (right) allows for real-time
  supervision and control of field-deployed devices. The rightmost panel shows
  an example dashboard, including sensor feeds and network status
  visualizations.}
\label{fig:2}
\end{figure*}

\section{The \textit{open storm} platform}

\textit{Open storm} provides a transparent and unified framework for sensing and
control of urban watersheds. To our knowledge, it is the only open-source,
end-to-end platform that combines real-time sensing, control and cloud services
for the purpose of water resources management. The project is designed to foster
engagement by lowering the technological barriers for stakeholders, decision
makers, and researchers. To this end, the \textit{open storm} framework is
accompanied by a body of reference material that aims to make it easy for
non-experts to deploy their own sensors and controllers. This living document,
available at \texttt{open-storm.org}, provides tutorials, documentation,
supported hardware, and case studies for end-to-end sensor network management.
In addition to documenting core features, this guide details the (literal)
\textit{nuts-and-bolts} of sensor network deployment, including information that
is typically not available in journal articles---such as mounting hardware,
assembly instructions and deployment techniques.

The \textit{open storm} framework can broadly be divided into three layers: hardware,
cloud services, and applications (Figure 2). The \textit{hardware} layer includes devices
that are deployed in the field---such as sensors for collecting raw data,
actuators for controlling water flows, microprocessors, and wireless
transmitters. The \textit{cloud services} layer includes processing utilities that
receive, store and process data, and interact with field-deployed devices
through user-defined applications. Finally, the \textit{application} layer defines how
users, algorithms, and real-time models interact with field-deployed devices.
This three-tier architecture allows for applications to be developed at a high
level, without the need for low-level firmware programming. Together, these
layers comprise a scalable framework that can easily be adapted to the needs of
a wide variety of users and applications.

\subsection{Hardware}

\subsubsection{The sensor node}

At its core, the \textit{open storm} hardware layer (Figure 1) is enabled by the sensor
node---a custom low-power embedded computer with wireless capabilities. The
sensor node collects measurements from attached sensors, transmits and receives
data from a remote server, and executes control actions. A microcontroller
(PSOC5-LP by Cypress Semiconductor) serves at the processing unit for the board.
This microcontroller is programmed with a simple operating system that schedules
the tasks to be executed, and interfaces with a series of device drivers that
control the behavior of attached sensors and actuators. The operating system is
designed to minimize power use and consists of a single routine which (i) wakes
the device from sleep mode, (ii) downloads pending instructions from the cloud
server, (iii) takes sensor readings and triggers actuators, (iv) transmits
sensor data to the server, and (v) puts the device back into sleep mode. The
sensor node spends the majority of its deployment in sleep mode, allowing it to
conserve battery power and remain in the field for an extended period of time.

The sensor node uses wireless telemetry to transmit and receive data from a
remote server. While internet connectivity can be achieved through a number of
wireless protocols, \textit{open storm} nodes currently use a cellular communications
protocol, which enables telemetry through 2G, 3G and 4G LTE cellular networks.
Cellular connectivity is implemented through the use of a cellular module (by
Telit), along with a small antenna for broadcasting the wireless signal.
Compared to other protocols (such as satellite or wi-fi), cellular
telemetry is especially suitable for urban and suburban environments due to (i)
consistent coverage, (ii) relatively low cost, and (iii) high data throughput.
At the time of writing, IoT cellular data plans can be purchased for under \$5
per month per node (1-10 MB), making it financially feasible for even small
research groups to maintain large-scale networks.

The sensor node is equipped with a power regulation subsystem to provide power
to the microcontroller and attached devices. The power supply system consists of
four components: (i) a battery, (ii) a solar panel, (iii) a charge controller,
and (iv) a voltage converter. The voltage converter permits the sensor node to
be powered across a range of 3-40V. While most sensor nodes are powered by a
3.7V Lithium Ion battery, 12V batteries can also be used for higher-voltage
sensors and actuators. The solar panel and solar charger are used to recharge
the battery, allowing the device to remain in the field without routine
maintenance. At the time of writing, many field-deployed sensor nodes have
reported data for over a year without loss of power.

Detailed technical information regarding the sensor node---including parts,
schematics and programming instructions---are available online at
\texttt{open-storm.org/node}. Excluding the cost of auxiliary sensors, the
sensor node can currently be assembled from off-the-shelf parts for a price of
approximately \$350 per node.

\subsubsection{Sensors and actuators}

The \textit{open storm} platform supports an extensive catalog of digital and analog
environmental sensors. Typical sensors include (i) ultrasonic and pressure-based
water level sensors, (ii) soil moisture sensors, (iii) tipping-bucket and
optical rain gages, (iv) automated grab samplers for assessing pollutant loads,
and (v) in-situ water quality sensors, including probes for dissolved oxygen,
pH, temperature, conductivity, dissolved solids, and oxidation-reduction
potential. While many sensors are known to work ``out of the box'', new sensors
can be quickly integrated by adding device drivers to the sensor node firmware.
Support for arbitrary sensors is provided by the microcontroller's
system-on-chip (SoC), which allows for analog and digital peripherals---like
analog-to-digital converters, multiplexers, and logic gates---to be generated
using programmable blocks in the device firmware. In addition to environmental
sensors, the sensor node also includes internal sensors that report device
health statistics, including battery voltage, cellular reception strength, and
connection attempts. These device health statistics help to diagnose network
issues, and can be used as inputs to remote trigger routines. Sensors can be
configured remotely using web services (see \textit{cloud services} section).
This capability allows users to turn sensors on or off, or to change the
sampling frequency of a sensor without reprogramming the device in the field.

The \textit{open storm} platform also supports an array of actuators that can be used to
move mechanical devices in the field. These devices are used to guide the
behavior of water systems in real-time, by controlling the flow of water in
ponds, channels and pipes. Butterfly valves are one common type of actuating device, and
are typically used to control discharge from storage elements such as retention
basins. Valves can be opened, closed, or configured across any number of
partially opened configurations to modulate flows. As with onboard sensors,
these devices are operated remotely using commands sent from a server. Control
signals can be specified manually, or through automated control algorithms.

Detailed technical information regarding supported sensors and actuators, along
with guides for integrating new devices are provided online at
\texttt{open-storm.org/sensors}.

\subsection{Cloud services}

While sensor nodes can function independently by storing data and making
decisions on a local level, integration with cloud services enables system-scale
supervision, configuration, and control of field-deployed devices. Like a
traditional SCADA system, the cloud services layer facilitates telemetry and
storage of sensor data, provides visualization capabilities, and enables remote
control of devices---either through manual input or through automated routines.
However, unlike a traditional SCADA system, the cloud services layer also allows
sensor nodes to communicate with a wide variety of user-defined web
applications---including advanced data visualization tools, control algorithms,
GIS software, external data ingesters, alert systems, and real-time hydrologic
models. By combining real-time supervision and control with domain-specific
tools, this architecture enables flexible system-scale control of water assets.

In brief, the cloud services layer performs the following core functions: (1)
stores and processes remotely-transmitted data, (2) simplifies management and
maintenance of field-deployed sensor nodes, and (3) enables integration with a
suite of real-time models, control algorithms, and visualizations. These
services are environment-agnostic, meaning that they can be deployed on a local
server or a virtual server in the cloud. In practice, however, current
\textit{open storm} projects are deployed on popular cloud services---such as
Amazon Elastic Compute Cloud (EC2)\cite{amazon_2017} or Microsoft
Azure\cite{microsoft_2017}---to ensure that computational resources flexibly
scale with demand. In the following section, we describe the
basic architecture, and present example applications that are
included with the \textit{open storm} platform.

The cloud services layer follows a simple design pattern, in which applications
communicate with sensor nodes through a central database. On the device side,
sensor nodes push sensor measurements to the database, and then query the
database to determine the latest desired control actions. On the server side,
applications query the latest sensor readings from the database, feed these
sensor readings into user-defined applications, and then write commands to the
database to control the behavior of field hardware remotely. This architecture
allows field-deployed sensors to be managed through a single endpoint, and also
allows new applications to be developed without modifying critical device
firmware.

The database serves dual purposes as both a storage engine for sensor data, and
as a communication layer between field-deployed sensors and web applications.
The primary purpose of the database is to store incoming measurements from
field-deployed sensors. Sensor nodes report measurements directly to the
database via a secure web connection---using the same protocol that one might
use to access web pages in a browser (HTTPS). The database address (URL) is
specified in the sensor node firmware, allowing the user to write data to an
endpoint of their choosing. In addition to storing sensor measurements, the
database also enables bidirectional communication between the node and
cloud-based applications by storing device configuration data, command signals,
and data from external sources. Server applications communicate with the sensor
node by writing commands to the database. These commands are then downloaded by
the sensor node on each wakeup cycle. For example, a real-time control
application might adjust outflow from a storage basin by writing a sequence of
valve positions to the database. At each sampling interval, the sensor node will
query the latest desired valve position and enact the appropriate control
action. This system enables bidirectional communication with field-deployed
sensor nodes without the need for complex middleware.

For its database backend, the \textit{open storm} project uses InfluxDB, a
time-series database that is optimized for high availability and throughput of
time-series data\cite{influxdata_2017}. Communications with the database backend
are secured through the use of basic authentication (i.e. a username and
password), as well as Transport Layer Security encryption (TLS/SSL). The use of
basic authentication prevents unauthorized parties from executing malicious
commands on the network, while the use of encryption prevents attackers from
intercepting sensitive data. Because applications communicate with the sensor
node through the database, this means that applications are secured
automatically against attackers as well. Altogether, this system comprises a
data backend that is secure, maintainable, and extensible.

\subsection{Applications}

The \textit{open storm} platform features a powerful application layer that
enables users to process and analyze data, build user interfaces, and control
sensor nodes remotely. Applications are implemented by creating a series of
subscriptions on the central database. These subscriptions perform one of three
actions: (i) \textit{read} from the database, (ii) \textit{write} new entries to
the database, and (iii) \textit{trigger} actions based on user-specified
conditions. While seemingly simple, this system allows for the development of a
wide range of applications. A data visualization platform, for instance, is
implemented by continuously querying sensor streams from the database;
similarly, automated control is implemented by writing a continuous stream of
commands. In the following section, we demonstrate the potential of the open
storm application platform by presenting example applications, including a data
visualization portal, a push alert system, adaptive control, and real-time
integration with hydrologic models.

\subsubsection{Network supervision and maintenance tools}

Much like a traditional SCADA system, the \textit{open storm} platform provides a
web-based graphical user interface for real-time visualization and device
configuration. Figure \ref{fig:2} shows an example dashboard, with time series
of cellular connection strength (top), radial gauges for monitoring battery
voltage (center), and real-time depth readings from two sensor nodes (bottom).
Time series visualizations are implemented using the Grafana analytics
platform\cite{grafana_2017}, which allows users to develop customized dashboards
that suit their individual needs. To facilitate remote configuration of sensor
nodes, \textit{open storm} also includes a web portal that allows users to change device
parameters (such as sampling frequency), control actuator behavior, and set
event triggers using a web browser.

\subsubsection{Automated alerts and adaptive control}

In addition to enabling manual supervision and control, \textit{open storm} also provides
a rich interface for triggering automatic actions based on user-specified
conditions. Push alerts are one common type of trigger event. Alerts can be used
to notify stakeholders of hazardous field conditions, such as flooding, or to
recommend control strategies to operators in real time. Alerts are also used to
notify the user about the health of the network---for instance, by sending push
warnings when node battery voltages drop below a threshold, or by emitting a
critical alert when data throughput ceases. These system health alerts allow
network outages to be promptly diagnosed and serviced. Alerts can be pushed to a
variety of endpoints, including email, text messages, or to social media
platforms such as Twitter and Slack\cite{twitter_2017, slack_2017}. The wide
variety of available push notification formats means that the \textit{open
  storm} platform is suited to handling both (i) confidential alerts for system
operators, and (ii) public emergency broadcasts.

In addition to the alert system, subscriptions are also used to implement
adaptive sampling and automatic control. Adaptive sampling allows the sampling
frequency of the node to be changed remotely in response to weather forecasts,
data anomalies, or manual user input\cite{wong_2016}. This in turn allows
hydrologically interesting events---such as storm events and dam releases---to
be measured at an enhanced resolution. To manipulate sampling frequencies in
response to changing weather conditions, for instance, weather forecasts are
first downloaded into the \textit{open storm} database using an external data
ingester. Next, the subscription service parses the incoming data. If the
service detects a probability of rain, the sampling frequency of a node is
increased. When no precipitation is anticipated, the sampling frequency is
decreased, allowing the node to conserve battery power. The same principle is
used to implement automated control. The subscription service can be configured
as a simple set-point or PID controller, for instance, by computing a control
signal based on an input data stream\footnote{An example script for a PID
  controller is included in the Supplementary Information document}. This
controller can in turn be used to optimize outflow from a retention pond, by
controlling the position of an outlet valve. More sophisticated control schemes
can be implemented by attaching the subscription service to an online model,
which optimizes control strategies over an entire stormwater network, achieving
system-level benefits. Examples include the MatSWMM and pySWMM software
packages\cite{Ria_o_Brice_o_2016, pyswmm_2017}, which are used to simulate
real-time control strategies for urban drainage networks.

Detailed information regarding cloud services and applications can be found on  
\texttt{open-storm.org/cloud}.


\begin{figure*}[!htb]
\centering
\includegraphics[width=\textwidth]{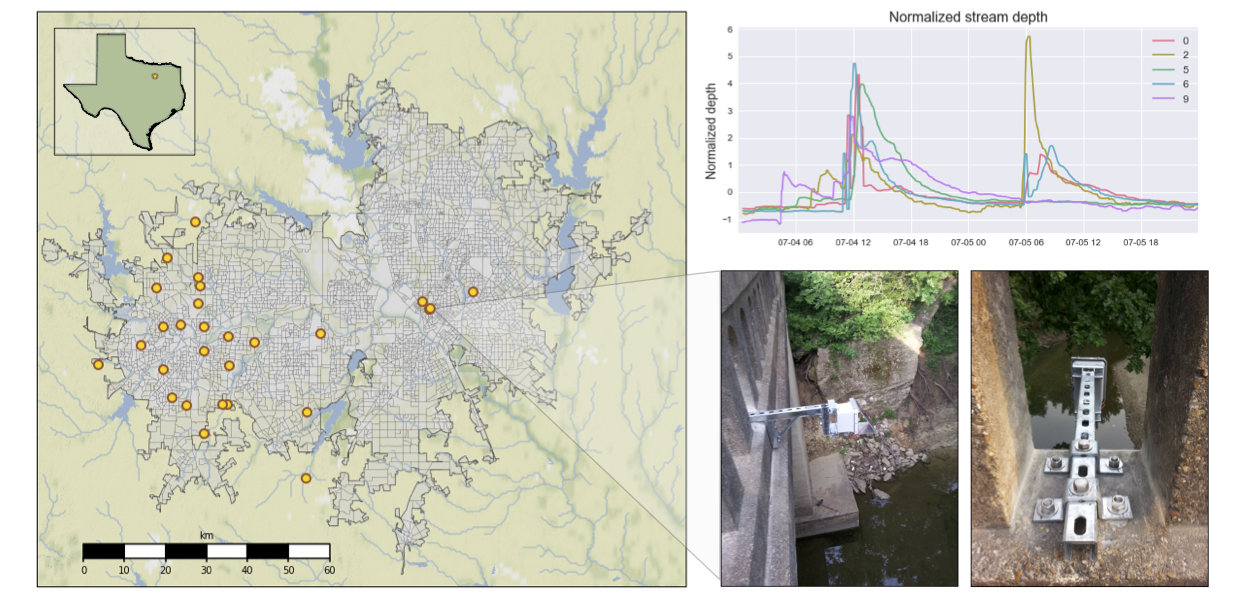}
\caption{Flood monitoring network in the Dallas--Fort Worth metroplex. The map
  (left) shows current and proposed sensor sites, while the detail photos
  (bottom-right) show an example bridge-mounted depth sensor node. Time series
  (top-right) show the response in stream depth to a series of storm events from
  August 5-6, 2016. From these stage hydrographs, it can be seen that the
  response varies widely even within a relatively small geographic area.}
\label{fig:3}
\end{figure*}

\section{Case Studies}

To demonstrate the capabilities of the \textit{open storm} platform, we present two
ongoing case studies. The first is a real-time flash flood warning
network for the Dallas--Fort Worth metroplex in Texas. This deployment detects
flash floods at the level of individual roadways, allowing targeted alerts for
motorists and improved routing of emergency services during storm events. The
second case study is a ``smart'' stormwater control network in the City of Ann
Arbor, Michigan. This deployment aims to improve water quality and mitigate
stormwater damage by adaptively timing releases from retention basins across an entire watershed.

\subsection{Case study 1: Flood monitoring}

Located in ``flash-flood alley'', the Dallas--Fort Worth (DFW) metroplex has
historically been one of the most flood-prone areas in the United States
\cite{sharif_2014}. Chronic flooding results in an average of 17 fatalities per
year in the state of Texas, with a majority of deaths arising from flash
floods \cite{sharif_2014}. Despite recent efforts to improve stormwater
management \cite{lee_leslie_2016}, lack of fine-scale runoff measurements inhibits
prediction and communication of flash flood risks. To address this problem, we
are using the \textit{open storm} platform to build a real-time flash flood monitoring
network. Drawing on the \textit{open storm} real-time alert system, this network aims to
improve disaster response by communicating flood risks to emergency managers in
real-time, and by generating targeted alerts that will allow motorists to safely
navigate around inundated roadways.

To date, urban flash flooding remains a poorly-understood phenomenon. There is
currently no model that is capable of generating reliable flash flood estimates
in urban areas \cite{Hapuarachchi_2011}. Modeling of urban flash floods is
complicated by an absence of natural flow paths and interaction of runoff with
man-made structures \cite{Hapuarachchi_2011}. However, lack of data at
appropriate spatial and temporal scales also presents a major challenge. For
reliable modeling of flash floods, Berne (2004) recommends using rainfall data
at a minimum spatial resolution of 500 meters\cite{Berne_2004}, while a recommended
temporal resolution of 1-15 minutes for rainfall is recommended by Smith (2007)
\cite{Smith_2007}. Existing rain gages and river stage monitors are often too
sparsely distributed to meet these requirements. Within the DFW metroplex, NWS
maintains 12 quality-controlled gages \cite{nws_2017}, while USGS provides precipitation
data at 30 sites \cite{usgs_2017}. This means that the current spatial
resolution of validated rain gages within the DFW metroplex is roughly 1 gage
per 600 km$^2$---too sparse for reliable prediction of flash floods. Likewise,
current river stage monitors for the DFW region are largely deployed along
mainstems of creeks and rivers with contributing areas ranging from 20 km$^2$ to
21,000 km$^2$ (and a median contributing area of 220 km$^2$). While these gages
provide excellent coverage of riverine flooding, they offer limited potential
for capturing flash floods.

To fill coverage gaps and enable real-time flash flood forecasting, we are
building a wide-area flood monitoring network that is specifically tailored to
monitoring flash floods over small-scale catchments (ranging from about 3 to 80
km$^2$ in size). Our approach is to leverage a large array of inexpensive depth
sensors to capture runoff response at the scale of individual roadways, creeks,
and culverts. By using inexpensive hardware, we are able to scale our network to
a size that would be infeasible with state-of-the-art stage monitoring stations
(such as those used by NOAA or USGS). At the time of writing, 40 sensor nodes
have been allocated and built for the DFW flood monitoring project, with over 15
nodes currently deployed and reporting. These 40 sensor nodes have been built
for a cost of \$20,000 USD---less than the cost as a single USGS gaging station
\cite{sheehan_2017}.\footnote{The installation cost for a USGS stage-discharge
  streamgaging station is roughly \$20,000, with an annual recurring cost of
  approximately \$16,000.}

Figure \ref{fig:3} presents an overview of the DFW flood monitoring network. The
left panel shows a map of the DFW metroplex, with current and proposed sensor
node locations. The bottom-left panel shows a detail of a typical sensor node
installation. Like most nodes in the network, this node is mounted to a bridge
deck with an ultrasonic depth sensor pointed at the stream surface below. The
sensor node records the depth to the water surface at a typical time interval of
3-15 minutes. The top-right plot shows a time series of stream depth during two
distinct storm events for a sample of nodes on the network. From this plot, it
can be seen that the runoff response varies widely between sensor locations,
even in a relatively concentrated geographic area. During the second event, for
instance, Node 2 (yellow) reports a large increase in discharge, while Node 9
(purple) reports no change in discharge. Comparison of the hydrographs with
NEXRAD\cite{Crum_Alberty_1993} radar data shows that the variability in stage is
largely explained by spatial variability in the rainfall fields\footnote{See
  https://github.com/open-storm/docs.open-storm.org/wiki/Case-study:-Flood-Monitoring-in-Dallas-Fort-Worth}.
This result confirms the need for increased spatial resolution in stream stage
measurements for flash flood monitoring.

The \textit{open storm} platform enables detection and communication of flood risks on
spatial and temporal scales appropriate for real-time disaster response and
control. Adaptive management of traffic during extreme weather events represents
one important application of this technology. The Dallas--Fort Worth flood
monitoring network could improve disaster response by communicating flood risks
to motorists in real-time, thereby allowing them to safely navigate around
flooded roadways. This is especially important given that in the US, roughly
74\% of fatalities from flooding are motor-vehicle related \cite{doocy_2013},
and in Texas, as much as 93\% of flood-related deaths result from walking or
driving into floodwaters \cite{sharif_2014}. Current alert systems are to a
large extent insensitive to spatial variability in flood response
\cite{smith_2009}. However, the \textit{open storm} framework enables targeted alerts
that can be integrated into existing mobile navigation apps. In a future that
may be characterized by autonomous vehicles and vehicle-to-infrastructure
communication \cite{jiang_2008}, this technology could one day be used to
adaptively route traffic during extreme weather events.


\begin{figure}[!htb]
\centering
\includegraphics[height=7cm]{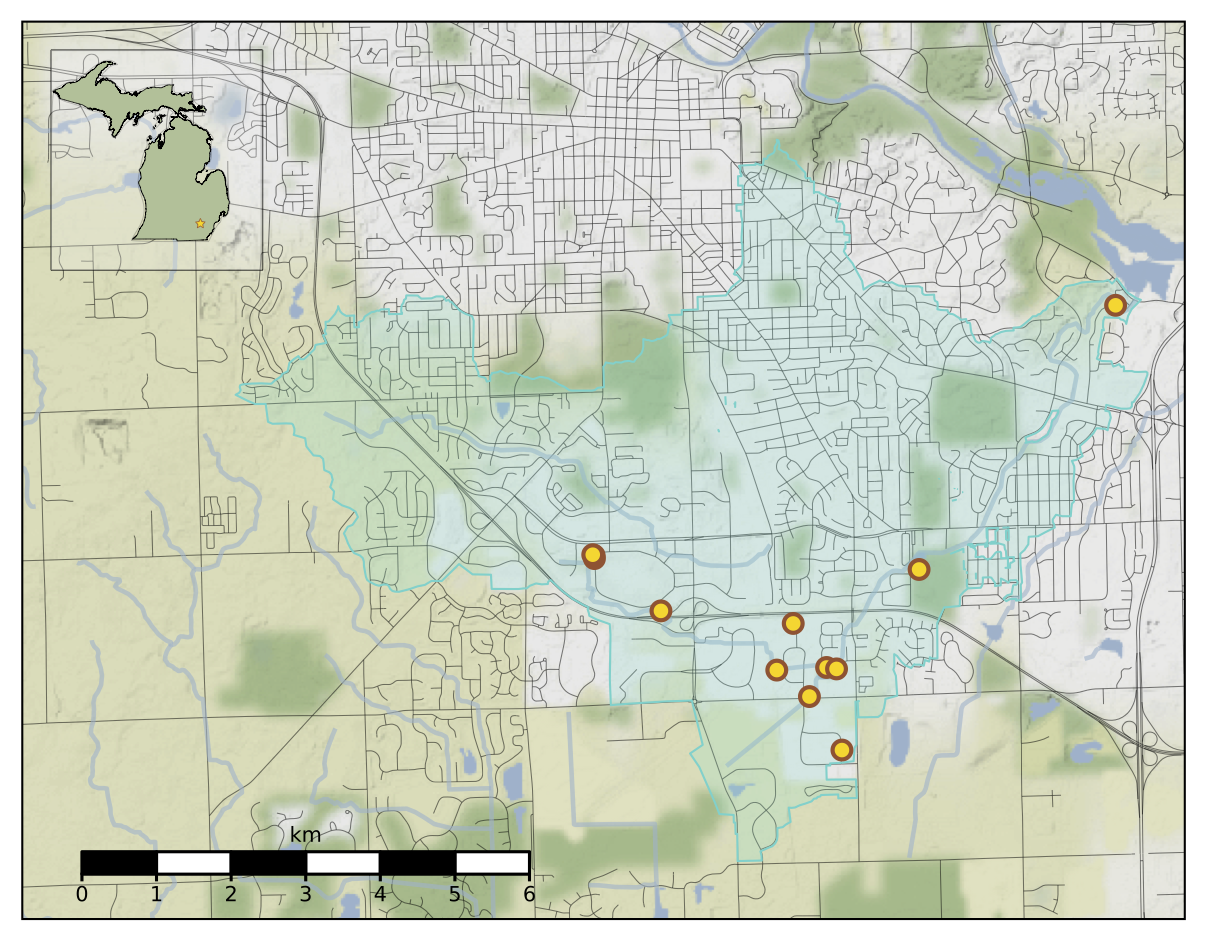}
\caption{Map of the Ann Arbor stormwater control network. Sensor nodes are
  concentrated within the impervious southern reach of the Malletts Creek
  watershed (blue). The outlet of the watershed drains into the Huron River
  mainstem (upper-right).}
\label{fig:4}
\end{figure}

\begin{figure*}[!htb]
\centering
\includegraphics[width=\textwidth]{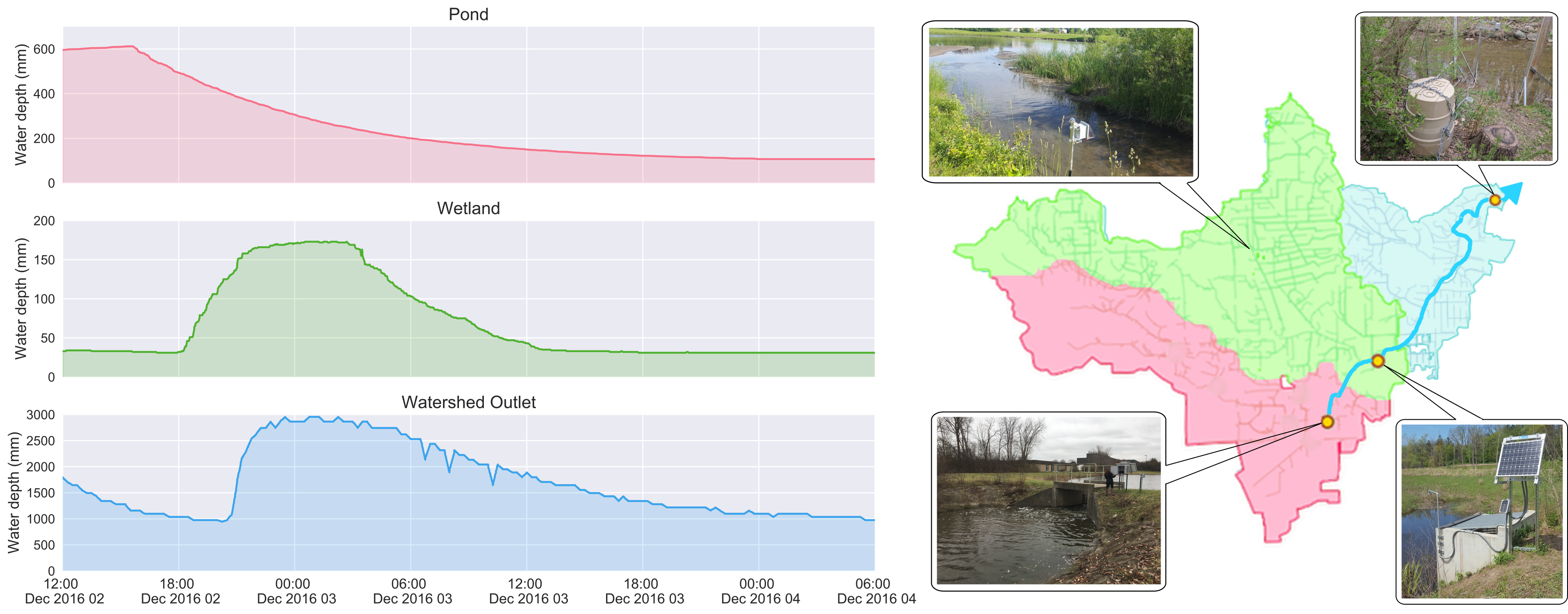}
\caption{Malletts Creek control experiment in Ann Arbor. The left panel shows time series of water depth from 12:00 pm on December 2 to 6:00 am on December 4, 2016. The right panel shows the location of the three sites in the watershed, with the partitioned contributing areas of each location corresponding to the colors of the time series plots.}
\label{fig:5}
\end{figure*}

\subsection{Case study 2: Controlling Watersheds}

As illustrated by the Dallas--Fort Worth flood-monitoring network, real-time
measurements can play a pivotal role in providing alerts to stakeholders and
improving our understanding of watershed dynamics. However, with the addition of
active control, it is possible to not only monitor adverse events, but also to
prevent them. The \textit{open storm} platform is capable of enacting control on a
watershed scale using distributed valve controllers, adaptive control schemes,
and cloud-hosted hydrologic models. Instead of building bigger stormwater
systems, operators may use real-time control to make better use of existing water
infrastructure, mitigate flooding, and decrease contaminant loads into sensitive
ecosystems.

The \textit{open storm} framework is presently being used to control an urban
watershed in the City of Ann Arbor, Michigan. The \textit{Malletts Creek} watershed---a
26.7 km$^2$ tributary of the Huron River---has traditionally served as a major
focal point in the city's strategy to combat flooding and reduce runoff-driven
water quality impairments \cite{hrwc_2011}. Given its proximity to the Great
Lakes, water resource managers have placed an emphasis on reducing nutrient
loads from urban runoff. A majority of the discharge in Malletts creek
originates from the predominantly impervious upstream (southwestern) reach of
the watershed, while a significant, but smaller portion of the discharge
originates from the central reach of the watershed. For this reason, local water
resource managers have constructed a number of flood-control basins in the
upstream segments of the catchment. It is these basins that are now modified to
allow for real-time control of the watershed.
  
The watershed is modified for control at two locations by retrofitting existing
basin outlets with remotely-operated valves (Figure \ref{fig:4}). The first
control point is a stormwater retention pond in the southern part of the
watershed (shown in red in Figure \ref{fig:5}). While originally designed as a
flow-through (detention) pond, the addition of two 30 cm diameter gate valves
allows for an additional 19 million liters of water to be
actively retained or released. The second control point is a smaller retention
pond, located in the central reach of the watershed (shown in green in
Figure \ref{fig:5}). This control site is retrofitted with a rugged 30 cm
diameter butterfly valve. The position of each valve is controlled via an
attached sensor node, which relays commands from a remote server. Each sensor
node is equipped with a pair of ultrasonic sensors: one to measure the water
depth at the pond, and one to measure the depth of the outflow stream. The
control sites operate entirely on 12V battery power, along with a solar panel to
recharge the battery during daylight hours. This configuration allows the
controller to remain in the field permanently, without the need for a dedicated
external electricity source.\footnote{With two people, installation at each site
  takes approximately one day. This includes time dedicated to mounting valves,
  sensors, and remotely-testing the equipment.}

In addition to the two control sites, the Ann Arbor network is also instrumented
with more than twenty sensor nodes that monitor system performance and
characterize real-time site conditions. Using a combination of ultrasonic depth
sensors, optical rain gages, and soil conductivity sensors, these nodes report
stream stage, soil moisture, soil temperature, and precipitation accumulation
approximately once every 10 minutes (with an increased resolution of 2-3 minutes
during storms). An additional set of nodes is deployed to measure water
quality---including dissolved oxygen, pH, temperature, oxidation reduction
potential, conductivity, temperature---as well as an automated grab sampler for
capturing contaminants of interest (such as heavy metals and microbes). These
nodes are deployed at the inlet and outlet of constructed wetlands to determine
how real-time control affects the removal of pollutants.

Measurements from the sensor network are validated using an external United
States Geological Survey flow measurement station (USGS station 4174518),
located at the watershed outlet. These federally-certified measurements are
available freely on the web, making them relatively easy to ingest into the
\textit{open storm} framework as an external data source. Furthermore, localized
weather forecasts are ingested from public forecasting services (darksky.net) to
provide daily, hourly, and minute-level forecasts to inform the control of each
site in the network\cite{darksky_2017}. These external data sources allow for
near-instant validation of sensor data, and provide a holistic ``snapshot'' of
system states.

We confirm the effectiveness of the control network through a simple experiment.
In this experiment, stormwater is retained at an upstream control site, then
released gradually to maximize sedimentation and reduce erosion downstream.
While it is known that the addition of control valves affords many localized
benefits---such as the ability to increase retention and capture sediments
\cite{gaborit_2013}---the goal of this experiment is to test the extent to which
control of individual sites can improve watershed-scale outcomes. The control
experiment takes place on a river reach that stretches across three sites: a
retention pond (upstream), a constructed wetland (center), and the watershed
outlet. Figure \ref{fig:5} (right) shows the three test sites within the
watershed, with the fractional contributing area of each site indicated by
color. In this system, runoff flows from the retention pond (red) to the
watershed outlet (blue) by way of an end-of-line constructed wetland (green)
designed to treat water, capture sediments, and limit downstream erosion.
Erosion, in particular, has been shown to be primary source of phosphorus in the
watershed \cite{wong_2016}, thus emphasizing the need to reduce flashy flows.
While the wetland serves a valuable purpose in improving water quality, it is
sized for relatively small events. Specifically, the basin is designed to hold
up to 57 million liters of stormwater but experiences as much as
760 million liters during a ten-year storm. Thus, it often overflows
during storms, meaning that treatment benefits are bypassed. To maximize
treatment capacity, a sensor node is placed into the wetland to measure the
local water level and determine the optimal time to release from the retention
pond upstream.

At the outset of the experiment, water is held in the upstream retention pond
following a storm on December 1, 2016. Residual discharge
from the original storm event can be observed as a falling hydrograph limb at
the USGS gaging station (blue) during the first 10 hours of the experiment
(Figure \ref{fig:5}). The sensor located at the wetland is used to determine the
time at which it is safe to release upstream flows without overflowing the
wetland (Figure \ref{fig:5}). Water is initially released from the pond at 4:00
pm on December 2, as indicated by a drop in the water level of the pond. Two
hours later, the water level in the wetland begins to rise due to the discharge
arriving from upstream. Finally, after another three hours, the discharge wave
reaches the outlet, where it is detected by the USGS flow station. Over the
course of the controlled release, the station registers roughly 19 million
liters of cumulative discharge.

The control experiment shows demonstrable improvements in system performance
compared to the uncontrolled case. While the water quality benefits will be
measured in the coming year, a number of likely benefits can be posited. As
measured, over 19 million liters were removed from the storm window and retained
in the basin following the storm event. The residence time of the water in the
pond increased by nearly 48 hours, increasing the potential for sedimentation
\cite{gaborit_2013}. The removal of stormwater flows also resulted in
attenuation of the downstream hydrograph. The peak flows at the watershed outlet
were measured to be 0.28 $m^3 / s$ during the storm, but would have been nearly
0.60 $m^3 / s$ had the valves in the basin not been closed. Based on prior
studies in the watershed---which showed that flows in the stream correlate
closely with suspended sediment concentrations---it can be estimated that the
flows from the basin were discharged at roughly 60 mg/L, rather than 110 mg/L,
thus nearly halving the concentration of suspended solids and total phosphorus
in the flows originating from the controlled basin\cite{wong_2016}. Moreover,
the controlled experiment enhanced the effective treatment capacity at the
wetland downstream, which would have overflowed during the storm, thus not
treating the flows from the upstream pond. As such, the simple addition of one
upstream valve provided additive benefits across a long chain of water assets,
demonstrating firsthand how system-level benefits can be achieved beyond the
scale of individual sites. While the water quality impacts of active control
deserve further assessment, this study opens the door for adaptive stormwater
control at the watershed scale. Rather than optimizing the performance of
isolated sites, the \textit{open storm} platform can be used to determine the
optimal control strategy for an entire watershed, then enact it in real-time.


\section{Conclusion}

\textit{Open storm} is an all-in-one, ``batteries included'' platform for monitoring and
managing urban water systems. Its emphasis on extensive configurability,
real-time response, and automated control make it an ideal choice for water
system managers and environmental researchers alike. While many open hardware
platforms exist, \textit{open storm} is the first open-source, end-to-end
platform that combines sensing, control and cloud computing in service of water
resources management. Aside from providing a technological blueprint, \textit{open storm}
addresses the real-world requirements that can be expected in water resources
applications, such as field-robustness, low-power operation and system-scale
coordination. The \textit{open storm} project has shown proven results in extending the
capabilities of existing stormwater systems: both by increasing the
spatiotemporal resolution of measurements, and by actively improving water
quality through real-time control. However, \textit{open storm} is not just a
platform---it's also a community of researchers, stakeholders and
decision-makers who are dedicated to realizing smarter water systems. To assist
in the dissemination and development of smart water systems, we are creating a
living document at \texttt{open-storm.org} in order to share standards,
reference materials, architectures, use cases, evaluation metrics, and other
helpful resources. We invite users to participate in this project by sharing
their experiences with designing, deploying and maintaining smart water systems.

\bibliography{main} 
\bibliographystyle{IEEEtran}

\end{document}